# Ubiquity of rotational symmetry breaking in superconducting films, from Fe(Te,Se)/Bi$_2$Te$_3$ to Nb, and the effect of measurement geometry

**Debarghya Mallick[1*]** | **Hee Taek Yi[1]** | **Xiaoyu Yuan[1]** | **Seongshik Oh[1]**

[1]136 Frelinghuysen Road, Piscataway, NJ - 08854, USA

**Correspondence**
Email: ohsean@physics.rutgers.edu, debarghyamallick121@gmail.com

**Present address**
[*]Oak Ridge National Laboratory, 1 Bethel Valley Road, Oak Ridge, TN 37830, USA

**Funding information**

FeTe$_{0.5}$Se$_{0.5}$/Bi$_2$Te$_3$ heterostructure is a promising new platform in the journey toward topological quantum computation, considering that first, FeTe$_{0.5}$Se$_{0.5}$ is itself known to be a topological superconductor (TSC) and second, the heterostructure has topological interface states that can be proximitized into TSC even if FTS fails to become TSC on its own. Here, we show that this system exhibits quasi-2D superconductivity, and utilizing the standard in-plane magneto-transport measurements, we discover two-fold anisotropy (a.k.a nematicity) in R$_{xx}$ and I$_c$ measurement, even though the system exhibits globally 12-fold symmetry. Then, we carried out similar measurements on a polycrystalline niobium (Nb) thin film, a well-known s-wave elemental superconductor, and found a similar two-fold symmetry even for this Nb system. This implies either that nematic behavior is ubiquitous or that the in-plane magneto-transport measurement scheme routinely used to detect nematicity is not a reliable method to probe nematicity. We show that the angle-dependent response of vortices in the superconducting regime to the magnetic Lorentz force is very likely the main cause behind the ubiquitous nematic behaviors of this measurement scheme. In other words, this mea-surement scheme is intrinsically two-fold, and is therefore not suitable to detect the nematicity. Accordingly, all the previous reports of nematicity based on similar measurement practices, reported on various samples, including thin films, bulk crystals, and exfoliated flakes, need to be reinterpreted.

## 1 | INTRODUCTION

The conventional picture of superconductivity was provided by the famous Bardeen-Cooper-Schrieffer (BCS) theory, where the electrons are paired by lattice vibration (phonon). Nevertheless, there have been other families of superconductors (SCs) that do not fall under this BCS category, where the pairing mechanism cannot be explained by the BCS theory. Iron-based superconductors (FeSCs) are one such family of SC, discovered in 2008, whose pairing mechanism is believed to be non-phononic [1]. Instead, spin fluctuations are considered to be the driving mechanism behind the unexpected superconductivity in FeSCs [2–6]. FeSCs are intriguing for several reasons. First, this family of SCs contains iron, which is a strong ferromagnet, and ferromagnets are antithetical to any SC. Second, they have a wide range of critical temperatures, ranging from a few kelvins to as high as 100 K: the only family of superconductors that exhibit higher critical temperatures ($T_c$ s) are cuprate SCs [7–9]. $FeTe_{1-x}Se_x$ is a member of this FeSC family and is of particular interest because it has a great potential to host the long-sought topological superconductivity and Majorana bound states when the Te content is about 50 %. There is a surge of research after the theoretical prediction and subsequent experimental verification of topological surface states (TSS) on $FeTe_{0.55}Se_{0.45}$ single crystal [10–21]. On the other hand, $Bi_2Te_3$ is a well-known topological insulator (TI) with TSS. Thus, $FeTe_{0.5}Se_{0.5}/Bi_2Te_3$, being a TSC/TI heterostructure, has an enhanced possibility, compared to just $FeTe_{0.5}Se_{0.5}$, to host the Majorana Fermion [22–29].

Nematic transition in superconductors is associated with the spontaneous breaking of rotational symmetry, keeping other symmetries preserved [30]. This transition is not a regular phonon-driven lattice deformation, but is driven by the electronic degrees of freedom. Since the discovery of FeSC in 2008, many SCs from this family have been investigated for the nematic transition, that is, breaking rotational symmetry from four-fold ($C_4$) to two-fold ($C_2$) (or tetragonal to orthorhombic) by various probes[31–33]. Because nematicity is in direct conflict with the s-wave symmetry, observation of nematicity is considered to be an important clue to uncover the symmetry of the superconducting order parameters [34, 35]. Accordingly, there have been intensive investigations on FeSCs to probe the nematicity, by scanning tunneling microscopy (STM) [22], angle-resolved photoemission spectroscopy (ARPES) [36], elastoresistance measurement [37], strain-dependent resistivity measurements [38], NMR [39], thermopower measurement [40], resonant inelastic X-ray scattering [41], electrical transport measurements [42] and so on.

One of the popular methods that have been used to probe the nematicity of a variety of superconducting systems is to measure magneto-transport properties while applying an in-plane magnetic field on a Hall-bar geometry, as shown in the cartoon of Fig. 3a. In this scheme, magneto transport properties such as $R_{xx}$ or $I_c$ or both are measured while the in-plane magnetic field is rotated 360 degrees. Using this measurement scheme, the emergence of rotation-symmetry breaking in or near the superconducting state while the rotation-symmetry is preserved in the normal state is considered a signature of nematicity in the superconducting order parameters. Following this protocol, signatures of nematic superconductivity have been reported on a variety of non-conventional superconducting platforms, including magic-angle graphene, thin films of Fe(Se, Te), $Sr_{1-x} La_x CuO_2$, thin film heterostructure of $YAlO_3/KTaO_3$, and single crystals of Sr doped $Bi_2Se_3$, $Ba_{0.5}K_{0.5}Fe_2As_2$, $RbV_3Sb_5$, $NbSe_2$, $CaSn_3$, $CeIr_3$, $Sn_4Au$, $Cu_{1.5}(PbSe)_5(Bi_2Se_3)_6$, Nb-doped $Bi_2Se_3$, $TaS_2$, Cu doped $Bi_2Se_3$ and so on [42–58].

Thus, such observation has been used to corroborate unconventional pairing mechanisms. One important point to note in this protocol is that it has been consensually assumed that if the same experimental scheme is followed on a conventional BCS superconductor, then such twofold rotational symmetry should not emerge. Nonetheless, however reasonable this assumption may look, in order to rule out the possibility of any artifact originating from the experimental setup, it is essential to carry out the same experiment on a well-known conventional (s-wave) superconductor and compare it with the system of interest. However, there is no report in the literature showing such comparison studies between conventional and non-conventional superconductors using the same measurement protocol. In other words, there has been a big hole in this wide-spread measurement protocol of probing nematic superconductivity. Here, we provide the first such comparison studies between a non-BCS and a BCS superconductor.

## 2 | RESULTS

We grew 15 nm thick $FeTe_{0.5}Se_{0.5}$(FTS) on 5 nm thick $Bi_2Te_3$(BT) on $Al_2O_3$ (0001) substrate as described in [24]. Although $FeTe_{0.5}Se_{0.5}$ and $Bi_2Te_3$ have different in-plane symmetries (4-fold vs. 6-fold), uniaxial lattice matching (Figure 1a) allows them to form epitaxial films with atomically sharp interfaces through hybrid symmetry epitaxy and the resulting $FeTe_{0.5}Se_{0.5}$ film on the $Bi_2Te_3$ layer exhibits globally 12-fold symmetric domain structures [24]. Then the heterostructure was patterned into a Hall bar with 15 $\mu m$ channel width using optical lithography and wet etching, as shown in the inset of Figure 1b. Resistance vs. temperature ($R_{xx}$ vs. T) data shows superconducting transition at around 11 K (Figure 1b) ($T_c$ has been estimated as the temperature, where the resistance becomes half of the normal resistance). Figures 1c and 1d show $R_{xx}$ vs. T measurements, both in a perpendicular and parallel magnetic field to the sample plane, up to 8.5 T.

Coherence length is an important parameter for SC to determine many of its properties, including the effective dimensionality of the superconductivity. We employed the phenomenological Ginzburg-Landau (G-L) theory [59]:

$$\mu_0 H_{c2}^{\perp}(T) = \frac{\Phi_0}{2\pi\xi_{GL}^2}\left(1 - \frac{T}{T_c}\right), \text{ or, } \xi_{GL} = \sqrt{\frac{\varphi_0}{2\pi\mu_0 \left|\frac{dH_{c2}^{\perp}(T)}{dT}\right| T_c}},$$

where $\Phi_0$, $\xi_{GL}$ and $\frac{dH_{c2}^{\perp}(T)}{dT}$ denote the flux quantum, coherence length and the slope of the curve to fit the perpendicular magnetic field ($B_\perp$) dependence of the critical temperature ($T_c$). From this, we find ($\xi_{GL}$) of our FTS/BT system to be around 1.6 nm. An important point to note is that this linear G-L equation is valid only near T $\approx T_c$ regime, so we fitted $B_\perp$ dependence of the $T_c$ in the narrow linear regime and extracted the slope at the point close to $T_c$ (Figure 1e). For an anisotropic system like FTS, $\xi_{GL}$ corresponds to the in-plane coherence length ($\xi_{ab}$). In order to measure the out-of-plane coherence length ($\xi_c$), we need to apply the magnetic field parallel to the sample plane, and the corresponding G-L equation should be: $\xi_c = \frac{\varphi_0}{2\pi\mu_0 \left|\frac{dH^{\parallel}(T)}{dT}\right| T_c \xi_{ab}}$, where $\frac{dH^{\parallel}(T)}{dT}$ is the slope near $T_c$ in the curve of the parallel magnetic field ($B_\parallel$) dependence of the critical temperature ($T_c$). This yields ($\xi_c$) of 0.62 nm. These values of $\xi_{ab}$ and $\xi_c$ are similar to what have been reported in the literature for other iron-based superconductors including Fe(Se,Te) [60–64].

Considering the layered structure of the FTS (and FTS/BT system) and that $\xi_c$ is somewhat smaller than $\xi_{ab}$, the system may effectively behave as a 2D system. If so, the superconducting transition is expected to follow Berezinskii-Kosterlitz-Thouless (BKT) mechanism. BKT transition describes how phase transition occurs in 2D through the binding and unbinding of vortex and anti-vortex pairs. In the superconducting regime, these vortices and anti-vortices are bound together and act like a single entity; hence, no energy dissipation. However, near and above the BKT transition temperature ($T_{BKT}$), due to thermal fluctuations, the vortex and anti-vortex pairs start to unbind and separate from each other, introducing energy dissipation, which, in turn, introduces finite resistance in the system. Now, to estimate $T_{BKT}$, we performed current-voltage (I-V) measurements at different temperatures from 2 K to 15 K (Figure 2a). It is evident from the figure that the transition from the superconductor to a normal state gets broadened with the temperature, and, in the completely normal state at 15 K, I-V follows Ohm's law. Below $T_{BKT}$, in the regime with the freely-moving vortices, I-V follows non-linear characteristics: $V \sim I^a(T)$ and, at $T = T_{BKT}$, $a(T_{BKT}) = 3$ [23, 45]. From the log-log plot of I-V (Figure 2b), we pulled out the value of $a$, which is the slope from the linear fitting. Figure 2c shows the variation of $a$ with the temperature and $T_{BKT}$ turns out to be 10.1 K. We can also employ Nelson-Kosterlitz relation: $R = R_0 exp[-b(\frac{T}{T_{BKT}} - 1)^{-1/2}]$ [65] or, reorganized as, $T - T_{BKT} \propto [\frac{d \ln R(T)}{dT}]^{-\frac{2}{3}}$ to extract the $T_{BKT}$. Figure 2d shows the extraction of $T_{BKT}$ to be 10.5 K from the Nelson-Kosterlitz relation. The fact that $T_{BKT}$ values extracted from the two different methods yield a similar value implies that the system is well described by the BKT transition with a well-defined $T_{BKT}$. Both the G-L analysis and the BKT transition analysis consistently suggest that the system behaves effectively as a quasi-2D system.

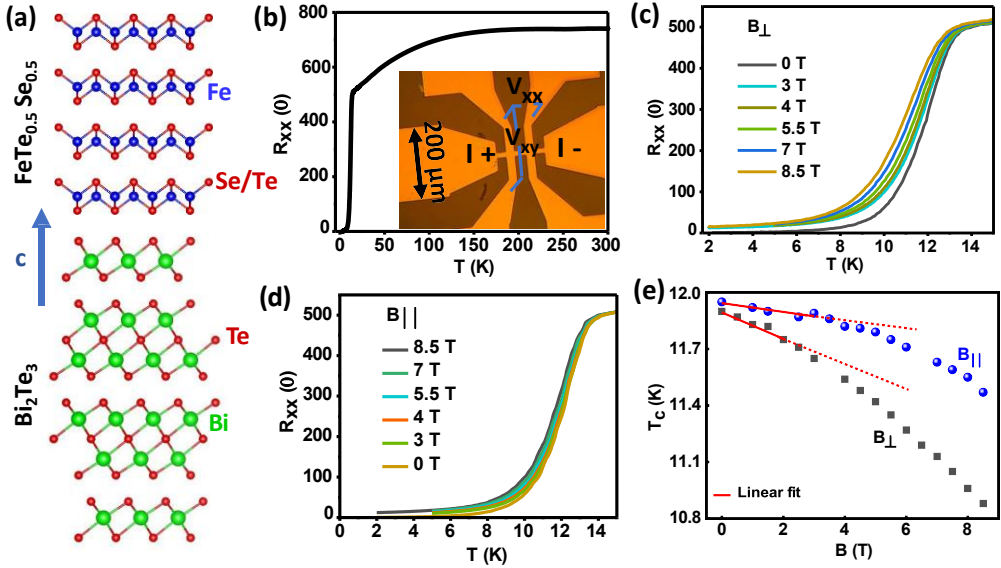

**FIGURE 1** Device image, schematics and resistance vs. temperature plots: (a) Schematics showing the lattice structures of the heterostructure $FeTe_{0.5}Se_{0.5}/Bi_2Te_3$. The four-fold symmetric $FeTe_{0.5}Se_{0.5}$ is grown on $Bi_2Te_3$ by hybrid symmetry epitaxy. (b) Resistance vs. temperature plot for the device from 2 K to 300 K. The inset shows the microscope image of the device: the central channel is 15 $\mu m$ wide. (c,d) Resistance vs. temperature plot for different perpendicular and parallel magnetic fields to the sample plane, respectively. (e) Critical temperature vs. magnetic field plots for both parallel and perpendicular alignment to the sample surface. The red lines are linear fittings to the points close to zero magnetic fields to extract the slope.

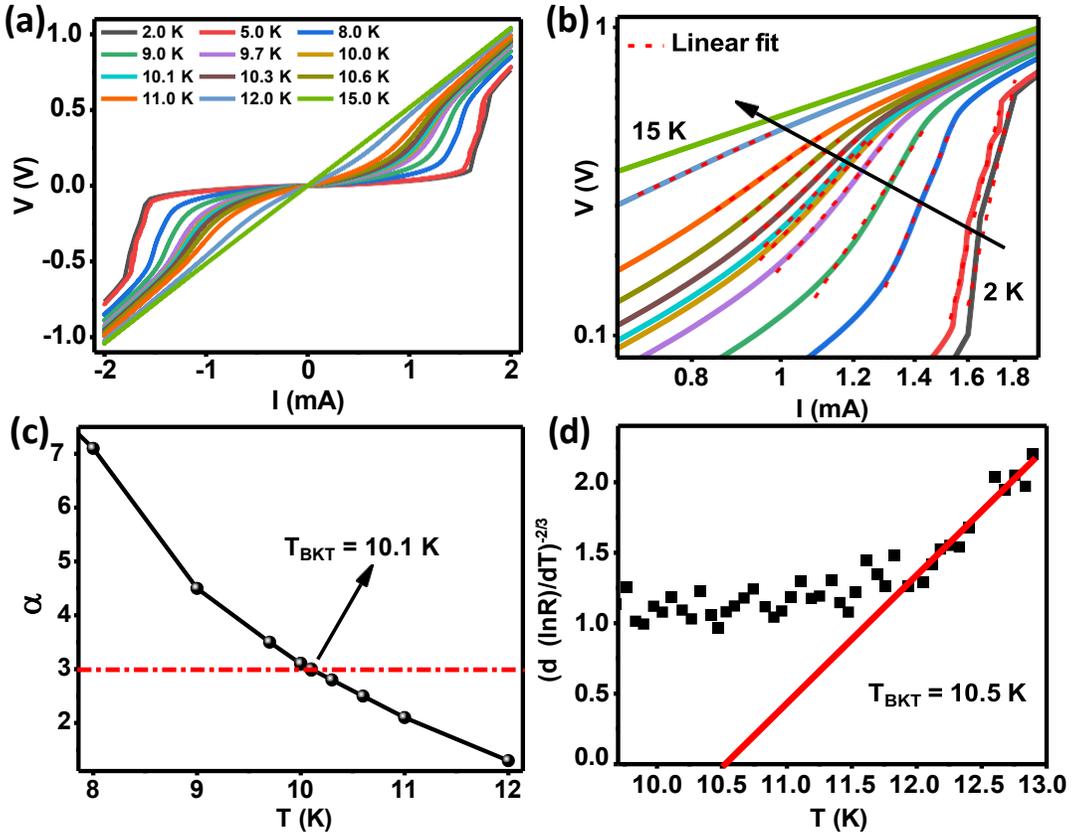

**FIGURE 2** Current-voltage characteristics: (a) Current-voltage plot for different temperatures from 2 K to 15 K, (b) log-log plot of (a). The linear fittings are shown in red lines to extract the slopes ($\alpha$) for each curve. (c) Slopes ($\alpha$) extracted from (b) are plotted with different temperatures. $\alpha = 3$ denotes the BKT transition temperature ($T_{BKT}$) and is shown as a dotted red line in the plot. $T_{BKT} = 10.1$ K (d) Reorganized resistance vs. temperature plot to fit with the Nelson-Kosterlitz formula yielding $T_{BKT} = 10.5$ K. This is to check the compatibility of $T_{BKT}$ extracted from the two separate methods.

Next, we probed the anisotropy of the critical current ($I_c$) and longitudinal resistance ($R_{xx}$) in the vortex phase of the SC, with the magnetic field applied parallel ($B_{\parallel}$) to the sample surface (Figure 3a). The critical current was measured from $0°$ to $350°$ with $B_{\parallel}$ being applied at 0 T, 3 T, 6 T and 8.5 T, while the temperature was fixed at 1.9 K. It is evident from Figure 3d that the critical current has two-fold anisotropy with $\theta$, defined in Figure 3a as the angle between the transverse direction to the channel and the direction of the magnetic field, $B_{\parallel}$. It is noteworthy that, when the magnetic field is zero, the anisotropy is almost negligible with $I_c$ nearly fixed at 1.75 mA, and, its anisotropy amplitude starts increasing gradually with the increase in the magnetic field. We have fitted the two-fold anisotropy using a sinusoidal function cos ($2\theta$ + c) where c is a constant term. Special care was taken to rule out any effect arising from the out-of-plane magnetic field component due to the inevitable slight misalignments. $I_c$ is maximum when $\theta$ is $90°$, that is, when $B_{\parallel}$ is parallel to the channel length, and it is minimum when it is perpendicular to the channel length.

Similarly, magnetoresistance also showed pronounced two-fold oscillations in the vortex regime of superconduc-tivity (Figure 3b,c). We have measured the modulation in magnetoresistance in two different ways. First, we fixed the temperature at 10 K and then varied $B_{\parallel}$ from 0 T to 8.5 T (Figure 3b). Second, we fixed $B_{\parallel}$ at 8.5 T and varied the temperature from 8 K to 15 K through $T_c$, which is around 11 K (Figure 3c). One important observation is that the modulation vanishes completely as the sample becomes normal beyond $T_c$. This directly indicates that the modulations are intrinsic to the SC phase. The comparison of Figure 3 b, c and Figure 3 d shows that the oscillations in $I_c$ and $R_{xx}$ are anticorrelated with each other. This is expected because they both signify the strength of the superconductivity and are inversely related to each other. Figure 3(e-g) are the polar plots of Figure 3(b-d). The same sinusoidal function was used to fit the two-fold modulation in resistance.

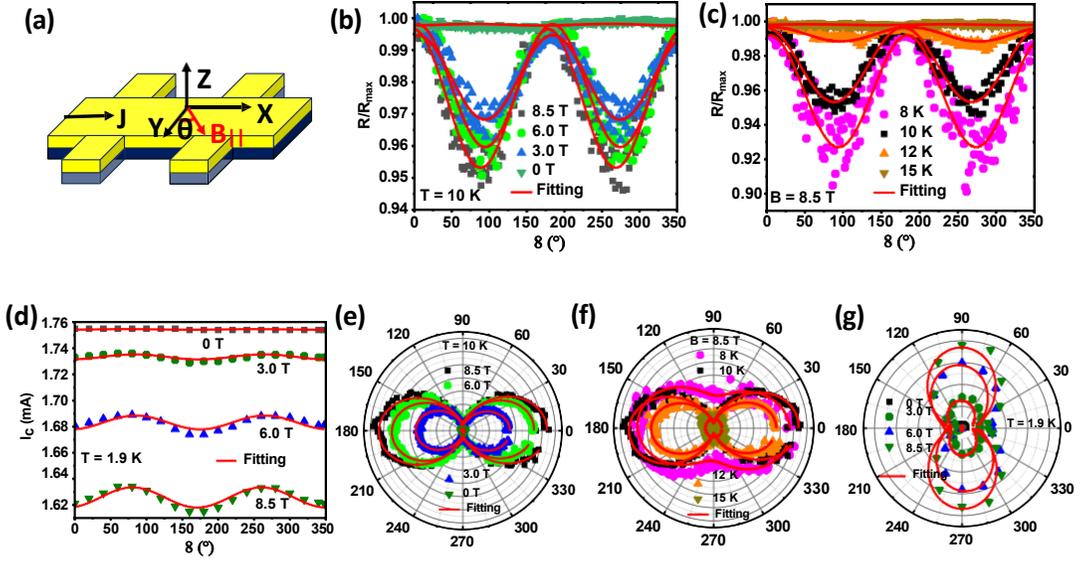

**FIGURE 3** Longitudinal resistance ($R_{xx}$) and Critical current ($I_c$) vs. the azimuthal angle ($\theta$) for FeTe$_{0.5}$Se$_{0.5}$/Bi$_2$Te$_3$: (a) Schematics of the device showing the coordinates and the current (J) direction in the channel. The azimuthal angle ($\theta$) is defined as the angle between $\hat{Y}$ and the magnetic field parallel to the sample plane ($B_{\parallel}$). (b) Magnetic field ($B_{\parallel}$) dependence of normalized longitudinal resistance ($R_{xx}$ vs. $\theta$). The temperature was kept at 10 K. (c)Temperature dependence of normalized longitudinal resistance ($R_{xx}$ vs. $\theta$). The magnetic field was kept at 8.5 T. Notably, the modulation completely disappears at 15 K, when the sample becomes normal. (d) The critical currents have been extracted from all the current-voltage plots performed at different azimuthal angles for four different parallel magnetic fields, namely, 0 T, 3 T, 6 T and 8.5 T and plotted against $\theta$. All the measurements are done at 1.9 K. (e-g) Polar plots of (b), (c), and (d), respectively. Red curves denote the fitting based on a two-fold sinusoidal function.

Nominally, this observation implies that the rotation symmetry of the system breaks into a two-fold symmetry when the system enters the superconducting state. Many previous reports attributed similar observations to the superconducting gap anisotropy caused by the possible mixture of s-wave and d-wave electrons because $I_c$ is directly related to the SC gap function [42–58]. However, none of these previous studies reported a control experiment comparing the results of these exotic superconductors with that of a conventional, s-wave superconductor. If the rotational symmetry breaking really is related to the symmetry of the superconductor order parameters, such nematic behavior should be absent in isotropic s-wave superconductors. Without such a control experiment, one cannot completely rule out the possibility of other trivial factors, such as the limitation of the experimental procedure itself.

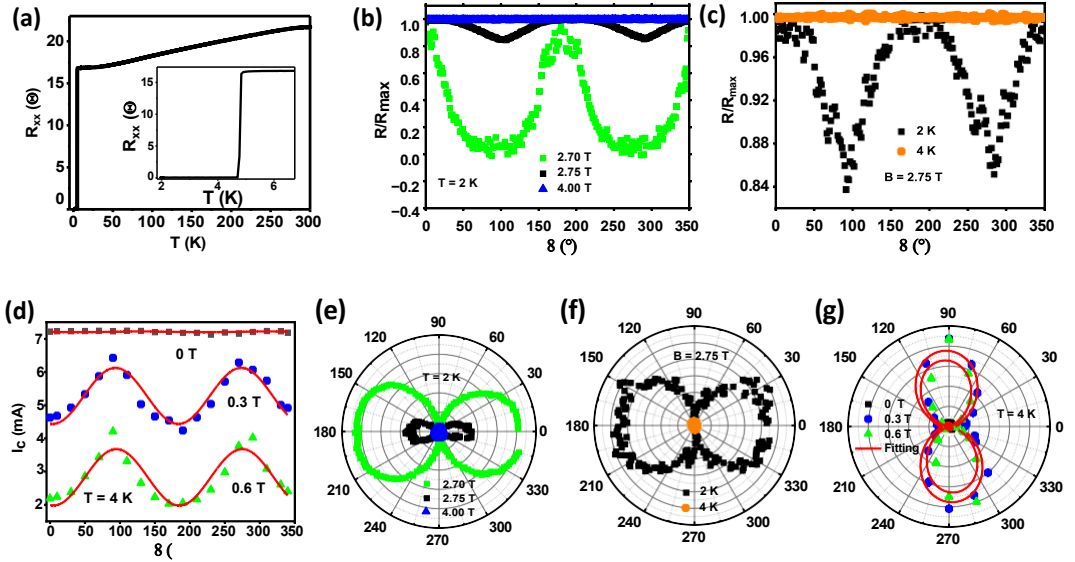

**FIGURE 4** Longitudinal resistance ($R_{xx}$) and Critical current ($I_c$) vs. the azimuthal angle ($\theta$) for a polycrystalline niobium thin film: (a) Resistance vs. temperature plot for the Nb device with 15 $\mu$m channel width. The zoomed-in segment of the plot at low temperatures is shown in the inset. (b) Magnetic field ($B_\parallel$) dependence of normalized longitudinal resistance ($R_{xx}$ vs. $\theta$). The temperature was kept at 2 K. (c) Temperature dependence of normalized longitudinal resistance ($R_{xx}$ vs. $\theta$). The magnetic field was kept at 2.75 T. Notably, the modulation completely disappears at 4 K when the sample becomes normal. (d) Critical current vs. $\theta$ plot shows the two-fold modulation. The measurements were done from 0° to 350° for three different in-plane magnetic fields: 0 T, 0.3 T, and 0.6 T. All the measurements were done at T = 4 K. (e-g) Polar plots of (b), (c) and (d), respectively. Red curves denote the fitting based on a two-fold sinusoidal function.

Accordingly, we chose niobium (Nb), a well-known elemental s-wave superconductor, as our control sample and fabricated a Hall bar structure on a sputter-deposited Nb thin film with an identical geometry to that of FTS/BT, and then carried out similar analysis. Figure 4a shows the $R_{xx}$ vs. T plot of the Nb film with $T_c$ of 4.7 K (although this $T_c$ is somewhat lower than that of high-quality Nb films, it is irrelevant for the purpose of the control experiment). Figure 4d shows the $I_c$ vs. $\theta$ plot, fitted with a sinusoidal function, for different in-plane magnetic fields (0 T, 0.3 T, and 0.6 T) at a fixed temperature, T = 4 K. We also carried out $R_{xx}$ vs. $\theta$ measurements by varying the magnetic field (Figure 4b) with the temperature fixed at 2 K and by varying the temperature (Figure 4c) with $B_\parallel$ fixed at 2.75 T. The polar plots for Figure 4(b-d) are shown in Figure 4(e-g), respectively. From Figure 4(b-g), it is evident that $R_{xx}$ shows two-fold symmetry in or near the superconducting state, while the two-fold symmetry completely vanishes when the system gets out of the superconducting state: this behavior is very similar to that of the FTS/BT system.

## 3 | DISCUSSIONS

Observation of the rotational-symmetry breaking for a simple elemental BCS superconductor, Nb, is unexpected and against the implicit assumption behind the well-established magnetotransport measurement protocol widely used in the literature. Being an elemental s-wave superconductor with an isotropic superconducting gap, a Nb film cannot have a two-fold-symmetric superconducting phase: This is particularly true because the Nb film is polycrystalline and therefore should be fully isotropic in their transport properties. Similarly, $FeTe_{0.5}Se_{0.5}/Bi_2Te_3$ film used in our work is also 12-fold symmetric due to the hybrid symmetry epitaxy [24], and therefore should also be highly symmetric for the transport properties in the plane. Nonetheless, both superconductors exhibit pronounced two-fold symmetry in the magneto-resistance data in the superconducting vortex phase, and the anisotropy completely vanishes once the sample becomes normal, either with temperature or with magnetic field.

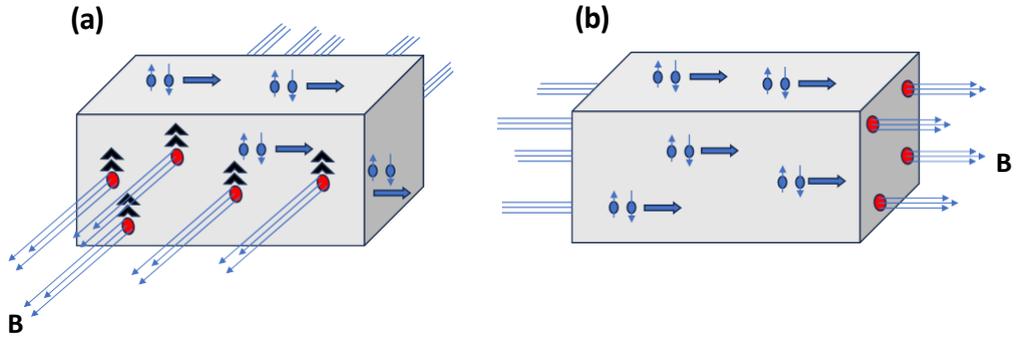

**FIGURE 5** Schematics for the Lorentz force acting on the vortices. In (a), the Lorentz force is along the Z-direction and is at its maximum because the magnetic field is perpendicular to the current flow direction, and the vortices start to move when the Lorentz force is larger than the vortex pinning force. In (b), there is no Lorentz force on the vortices because the magnetic field is along the current flow direction. Magnetic field lines, vortices (red dots), Cooper pairs, current flow (thick blue) and the Lorentz force (black arrows) are shown in the figures.

It is important to note that the geometry of the in-plane magneto-transport measurement scheme is intrinsically two-fold, which is not much discussed in the literature, though. We can see this by considering the Lorentz force felt by charged carriers due to the applied magnetic field (**B**), which is given by $\mathbf{F} \sim \mathbf{I} \times \mathbf{B}$, where **I** and **B** are the current flowing through the sample and the magnetic field, respectively. It is evident that while the Lorentz force is zero when the magnetic field is applied in parallel to the current flow direction, its magnitude grows and reaches a maximum when the magnetic field is perpendicular to it. In other words, the magnitude of the Lorentz force felt by the charged carriers is intrinsically two-fold against the direction of the in-plane magnetic field. Then, a natural question arises as to why this two-fold symmetry is observed only when the sample is in the superconducting transition regime but absent when the sample is in the normal state.

We can explain this as follows. When **B** is in the plane but perpendicular to **I,** the Lorentz force is in the out-of-plane direction. Then charges of opposite types will accumulate on the top and bottom planes like a parallel capacitor of the thin film structure, which will generate an electric field, until it exactly cancels the Lorentz force caused by the magnetic field. Accordingly, in the normal state, the charge carriers flow through the thin film structure as if there is no magnetic field applied, regardless of the magnetic field direction, as far as the magnetic field is in the plane. This explains why we do not see the two-fold symmetry when the sample is in the normal state.

Now, if the film is fully in the superconducting state, because the resistance is completely zero, there is no anisotropy either. However, when the applied in-plane magnetic field is larger than a critical field, vortices (nanoscopic non - superconducting region) are formed surrounded by superconducting fluids as shown in Figure 5. In such a case, even though superconducting fluids repel the magnetic fields, the vortices do feel the Lorentz force, and if it is larger than the pinning force, the vortices will start to move. Then, the movement of these vortices causes dissipation and in turn, this kills the superconductivity [66–68] . The important point here is that because the magnitude of the Lorentz force is two-fold (maximum when I $\perp$ B and minimum when I || B), the resistance or critical currents in the superconducting transition regime, where superconducting fluids and vortices coexist, should also be two-fold, regardless of the underlying symmetry of the superconducting order parameter. According to this analysis, the two-fold symmetric in-plane magneto-transport measurements should be a natural result of the measurement geometry. It is also important to note that all the in-plane magneto-transport measurements in the literature relying on a similar geometry report two-fold symmetry regardless of the superconducting materials, and they are almost always interpreted as a signature of unconventional (non-s-wave) superconductivity. The very observation of the two-fold symmetry from a fully isotropic polycrystalline Nb film, whose transport properties cannot be anything but isotropic, strongly supports that two-fold symmetry originates from the very measurement geometry as discussed above, rather than the unconventional gap symmetry of the superconducting order parameter. However, this observation does not necessarily imply that the two-fold symmetry should originate entirely from the measurement geometry. Suppose the superconductor has its own anisotropic gap symmetry. In that case, the measured anisotropy should be a combined effect of both gap symmetry and the two-fold measurement symmetry, which could be the case for some of the result in the literature. Regardless, this study shows that the widely used in-plane magneto-transport measurement is not a reliable method to determine the nematicity of the superconducting order parameter. In order to disentangle the effect of the measurement geometry and properly probe rotational symmetry breaking, it is imperative to use an isotropic device structure such as the Corbino geometry. Even in such a case, in order to rule out the possibility of any other artifact, it should still be necessary to test it on an isotropic trivial superconductor such as Nb before making bold claims on any exotic superconductors. This should be an interesting subject for future studies.

In summary, we carried out the first comparison study between an elemental BCS superconductor (Nb) and an unconventional superconductor (FeTe$_{0.5}$Se$_{0.5}$/Bi$_2$Te$_3$) in the context of detecting rotational symmetry breaking in superconducting states, utilizing the widely used in-plane magnetotransport measurement scheme. From this study, we showed that both systems exhibit similar two-fold anisotropy in magnetoresistance and critical current on Hall bar devices. For both devices, the superconductivity strength is maximum when the magnetic field is along the channel length and minimum when the magnetic field is transverse to the channel, and the two-fold modulation vanishes once the sample becomes normal. Similar results for the two very different systems - one being a TSC/TI heterostructure with supposedly s$_\pm$ pairing symmetry and the other being an elemental s-wave superconductor - indicate that one should be careful claiming unconventional superconductivity based on just two-fold symmetric transport signatures of superconductivity. Considering that other groups in the literature have widely used this measurement scheme, this observation raises an important question on whether the two-fold symmetric transport signatures are a common phenomenon regardless of the pairing symmetry or whether this measurement protocol is unsuitable for detecting the nematic phase. Our analysis shows that this measurement scheme is fundamentally unsuitable to detect the nematic phase, because the impact of the Lorentz force on the vortices in the superconducting regime is intrinsically two-fold in this measurement geometry. Considering the scientific significance of nematicity in superconducting research, all the previous reports of nematic phases relying on similar measurement schemes will have to be revisited and reinterpreted.

# 4 | EXPERIMENTAL SECTION

## 4.1 | Thin films growth

The thin film heterostructure, $FeTe_{0.5}Se_{0.5}/Bi_2Te_3$ was grown on $Al_2O_3(0001)$ substrate by molecular beam epitaxy (MBE) in a vacuum chamber with a base pressure of $10^{-10}$ Torr. The growth details can be found in [24] and the supplementary information therein. The niobium thin films (40 nm) were sputtered at room temperature on $Al_2O_3(0001)$ substrate using DC magnetron sputtering on a 2" diameter Nb target. Substrates were cleaned ex-situ by 5 min of exposure to UV-generated ozone before mounting in the sputtering chamber, which was subsequently pumped down to $10^{-7}$ Torr, before 2 mTorr of argon was introduced into the chamber for sputtering: the power to generate the plasma was kept at 200 W.

## 4.2 | Device fabrication

We utilized UV photolithography using a Karl-Suss mask aligner MJB3, followed by wet etching to pattern a 15 $\mu m$ width Hall bar for both films. The wet etchants used for etching the FTS/BT film was Cr-etchant from Transene (which is a mixture of $HNO_3$ and $NH_4NO_3$), and that for the Nb thin film was Nb-etchant from Transene (which is a mixture of $HNO_3$ and HF). The samples were post-baked with an optimized recipe in order to minimize the undercut associated with the wet etching. After the device fabrication, the samples were immediately coated with a PMMA layer to prevent degradation from the atmosphere.

## 4.3 | Electrical transport

All the devices were measured in a Quantum Design Physical Properties Measurement System (PPMS), with a base temperature of 2 K and a maximum magnetic field of 9 T. For the anisotropy measurements, the magnetic field was kept fixed, and the sample was mounted on a puck that could rotate azimuthally. We used indium wires to make the electric contact from the sample to the puck.

# 5 | ACKNOWLEDGEMENT

This work is supported by Army Research Office's MURI W911NF2020166 and the center for Quantum Materials Synthesis (cQMS), funded by the Gordon and Betty Moore Foundation's EPiQS initiative through grant GBMF10104. The authors would like to thank Saikat Banerjee, Phanibhusan Singha Mahapatra, Pratap Raychaudhuri and Lara Benfetto for valuable discussions.